\documentclass[12pt]{article}
\usepackage{graphicx}

\def\pbnr{}
\def\speaker{Christine Davies}
\def\onbehalfof{}
\def\title{Precise Determination of the Charm Quark Mass}
\def\affiliation{School of Physics and Astronomy\\
The University of Glasgow, Glasgow, UK}
\def\support{This work was supported by STFC, the Royal Society and the 
the Wolfson Foundation}
\newcommand\pubnumber{\pbnr}
\newcommand\pubdate{\today}
\def\Title#1{\begin{center} {\Large #1 } \end{center}}
\def\Author#1{\begin{center}{ \sc #1} \end{center}}

\newcommand{\OnBehalf}[1]{\sbox0{#1}\ifdim\wd0=0pt
        {}% if #1 is empty
	\else
	{\\on behalf of #1}% if #1 is not empty
	\fi}
\newcommand{\SupportedBy}[1]{\sbox0{#1}\ifdim\wd0=0pt
        {}% if #1 is empty
	\else
	{\footnote{#1}}% if #1 is not empty
	\fi}
\def\Address#1{\begin{center}{ \it #1} \end{center}}

\newcommand\pubblock{\includegraphics[width=5cm]{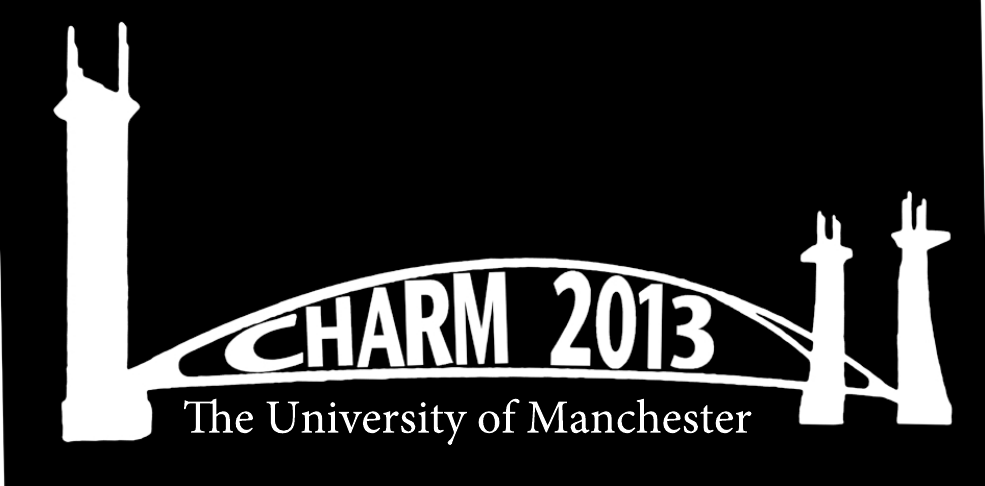}\hfill{\begin{tabular}{l} \pubnumber\\
         \pubdate  \end{tabular}}}
\newenvironment{Abstract}{\begin{quotation}  }{\end{quotation}}
\newenvironment{Presented}{\begin{quotation} \begin{center} 
             PRESENTED AT\end{center}\bigskip 
      \begin{center}\begin{large}}{\end{large}\end{center} \end{quotation}}
\def\Acknowledgements{\bigskip  \bigskip \begin{center} \begin{large}
             \bf ACKNOWLEDGEMENTS \end{large}\end{center}}
\def\venue{The 6$^{th}$ International Workshop on Charm Physics\\
(CHARM 2013)\\
Manchester, UK,  31 August -- 4 September, 2013}
\def\beq{\begin{equation}}
\def\eeq#1{\label{#1}\end{equation}}
\def\eeqn{\end{equation}}

\def\beqa{\begin{eqnarray}}
\def\eeqa#1{\label{#1}\end{eqnarray}}
\def\eeqan{\end{eqnarray}}

\let\bar=\overbar

\def\Dslash{\not{\hbox{\kern-4pt $D$}}}
\def\dslash{\not{\hbox{\kern-2pt $\del$}}}

\def\msb{{\bar{\ssstyle M \kern -1pt S}}}

\begin{document}
\begin{titlepage}
\pubblock

\vfill
\Title{\title}
\vfill
\Author{\speaker\SupportedBy{\support}\OnBehalf{\onbehalfof}}
\Address{\affiliation}
\vfill
\begin{Abstract}
%%%%%%%%%%%%%%%%%%%%%%%%%%%%%%%%%%%%%%%%%%%%%%%%%%%%%%%%%%%%%%%%%%%%%%%%%%%
The determination of the charm quark mass is now possible 
to 1\% from QCD, with lattice QCD pushing the error down 
below 1\%. I will describe the ingredients 
of this approach and how it can achieve this accuracy. 
Results for quark mass ratios, $m_c/m_s$ and $m_b/m_c$, 
can also be determined to 1\% from lattice QCD, allowing 
accuracy for the heavy quark masses to be leveraged 
into the light quark sector. 
I will discuss the prospects for, and importance of, improving 
results in future calculations.
%%%%%%%%%%%%%%%%%%%%%%%%%%%%%%%%%%%%%%%%%%%%%%%%%%%%%%%%%%%%%%%%%%%%%%%%%%%
%A big thank you to everyone who made CHARM 2013 a success!
\end{Abstract}
\vfill
\begin{Presented}
\venue
\end{Presented}
\vfill
\end{titlepage}
\def\thefootnote{\fnsymbol{footnote}}
\setcounter{footnote}{0}
%

%%%%%%%%%%%%%%%%%%%%%%%%%%%%%%%%%%%%%%%%%%%%%%%%%%%%%%%%%%%%%%%%%%%%%%%%%%%
%  WHAT FOLLOWS IS YOUR TEXT
%%%%%%%%%%%%%%%%%%%%%%%%%%%%%%%%%%%%%%%%%%%%%%%%%%%%%%%%%%%%%%%%%%%%%%%%%%%
\section{Introduction}
\label{sec:intro}
Quark masses are important parameters of the Standard Model but 
cannot be obtained directly from experiment because quarks are 
never seen as free particles. Instead they must be inferred from 
experimental results for hadrons. The accuracy of the determination 
of quark masses is a topical issue because of the need to test 
the couplings to quarks of the newly discovered Higgs boson~\cite{lhcwg1, lhcwg2, snowmasswg}. The 
Standard Model rate for decay of a Higgs to $c\overline{c}$ or $b\overline{b}$  
is sensitive to the charm/bottom quark mass. 

The quark mass parameter in 
the QCD Lagrangian is a well-defined quark mass but it is
scheme- and scale-dependent (i.e. it `runs'). Lattice QCD has a 
clear advantage here when determining quark masses, because 
the calculations start from the QCD Lagrangian and the parameters 
of that Lagrangian are readily tuned. To do this, quark mass parameters are 
chosen, at a given value of the lattice spacing, to reproduce
the experimental result for the mass of a hadron containing 
that quark. This gives the quark mass in the lattice scheme 
very accurately. However, most calculations (such as those for 
Higgs decay) need quark masses 
in a continuum renormalisation scheme such as $\overline{MS}$. 
A key source of error is then the conversion from the lattice 
quark mass to the $\overline{MS}$ scheme. 

Continuum methods for 
determining the quark mass rely on evaluating a quantity 
from experiment that can also be calculated accurately 
in QCD perturbation theory in terms of, say, the $\overline{MS}$ 
quark mass. As discussed below, accurate values
for $c$ and $b$ masses can be obtained in this way 
using experimental results derived 
from $\sigma(e^+e^- \rightarrow \mathrm{hadrons})$~\cite{karlsruhe1, karlsruhe2}. 
Very similar methods can be used with lattice QCD 
results~\cite{hpqcdkarlsruhe, hpqcd10} 
effectively to convert the lattice quark mass to the $\overline{MS}$ 
scheme, and it is these methods that give the most accurate 
results from lattice QCD also.    

I will describe both methods and their results in Section~\ref{sec:currcurr} 
but first give a brief introduction to lattice QCD. 

\section{Lattice QCD Calculations}
\label{sec:lattice}
Lattice QCD calculations proceed by a standard recipe~\cite{DeGrandDeTar} which starts 
with setting up a 4-d space-time volume, discretised into a 
set of points with lattice spacing, $a$. Configurations 
of gluon fields (one SU(3) matrix for 
every link joining two points on the lattice)  
are generated by Monte Carlo methods according to the 
probability distribution required in the QCD Feynman Path 
Integral. This probability distribution is $\exp(-S_{QCD})$ where 
$S_{QCD}$ is the sum over the configuration of the Lagrangian 
of QCD. The probability distribution is for the gluon fields but, in 
modern lattice QCD calculations, it includes the effect of sea quarks 
that are generated in the `soup' of particles that make up 
the QCD vacuum. The parameters of QCD enter in specifying the 
QCD Lagrangian. These are the bare coupling constant and the 
quark masses. It is important to realise that the lattice spacing 
is {\it not} specified at this point - it must be determined from 
calculations performed on these configurations. 

%%%%%%%%%%%%%%%%%%%%%%%%%%%%%%%%%%%%%%%%%%%%%%%%%%%%%%%%%%%%%%%%%%%%%%%%%
%%
%%   use this format to include an .eps figure into your paper
%%
\begin{figure}[t]
\centering
\includegraphics[width=6.0cm]{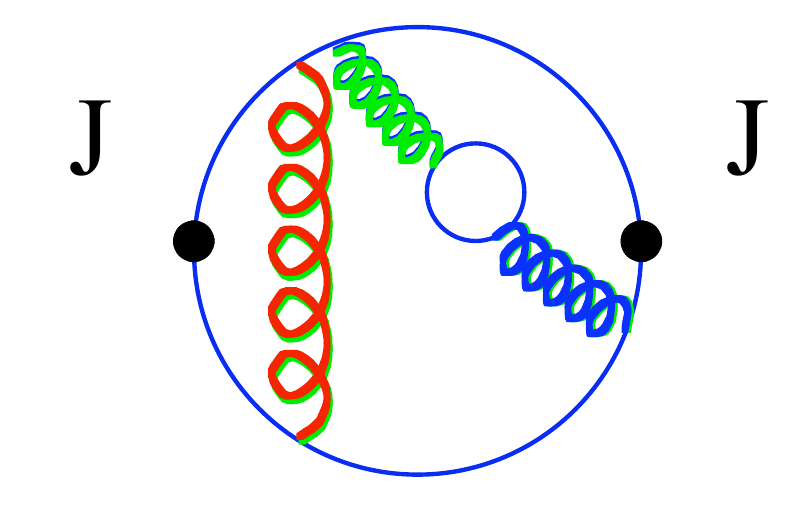}\hspace{1.0cm}
\includegraphics[width=8.0cm]{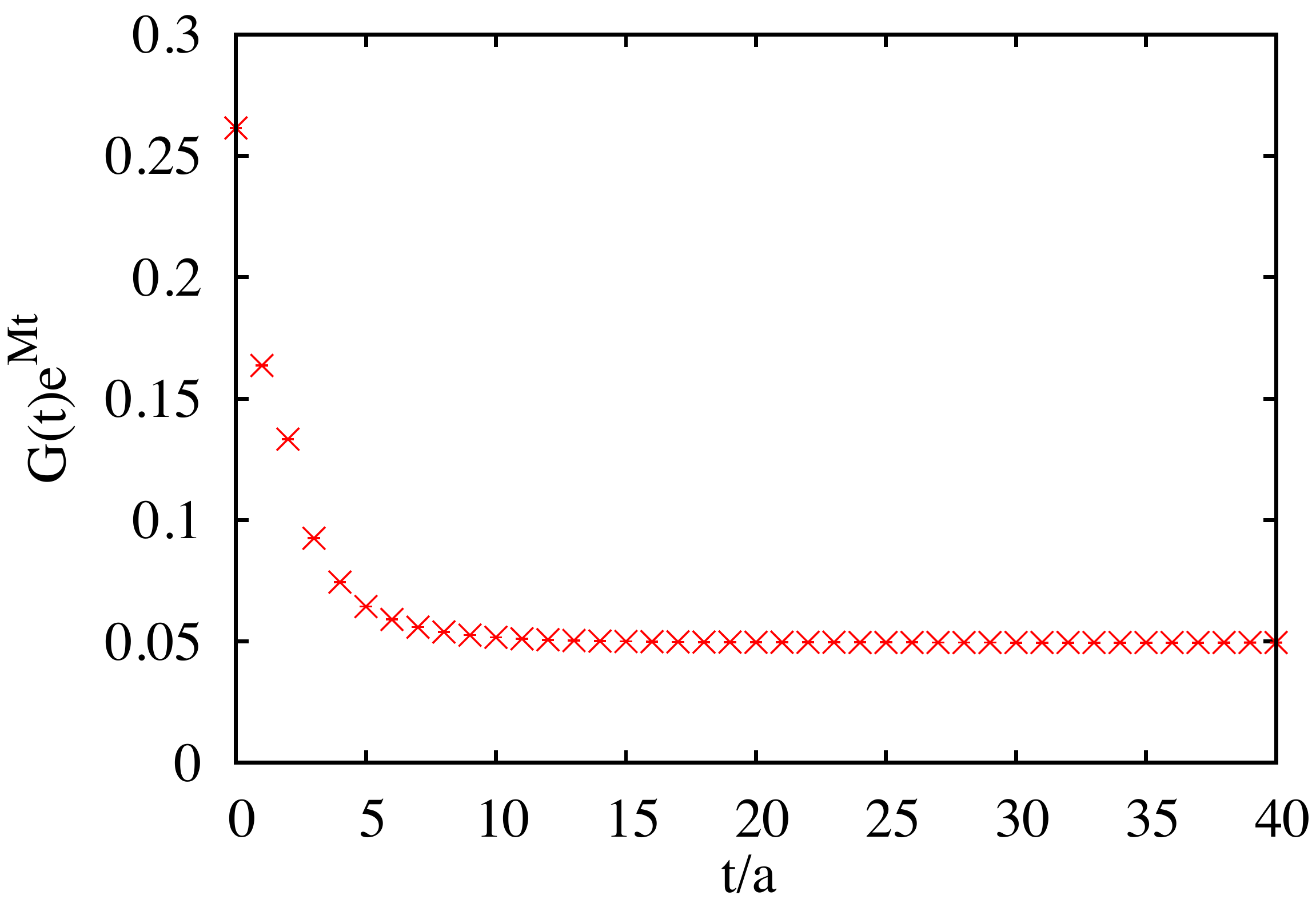}
\caption{Left: a $c\overline{c}$ meson correlation 
function in QCD and also the charm quark vacuum polarisation. Charm quark propagators 
connect the two currents, $J$. Right: The correlation 
function for a $c\overline{c}$ pseudoscalar meson multiplied by $e^{Mt}$ 
(where $M$ is the fitted ground-state mass) 
and plotted against time $t/a$ in lattice units. The ground-state clearly 
dominates the correlation 
function at large $t$. The statistical errors from the lattice calculation 
are shown, but are so small as to be barely visible. }
\label{fig:masses}
\end{figure}
%%%%%%%%%%%%%%%%%%%%%%%%%%%%%%%%%%%%%%%%%%%%%%%%%%%%%%%%%%%%%%%%%%%%%%%%%%%

Once sets of gluon field configurations have been generated, we 
can calculate quark propagators on them by solving the Dirac equation. 
In this equation the gluon field appears in the covariant derivative 
term and the quark mass is a parameter. Combining a quark and 
antiquark propagator together (making sure the colours match 
at both ends and the spins are combined appropriately) 
makes a meson correlation function. This is the amplitude to 
create a meson at one point and destroy it at some other 
point. Averaging the meson correlation functions obtained 
over all the gluon field configurations generated 
gives us a Monte Carlo estimate of the result for 
this amplitude from the QCD Feynman Path Integral. 
The meson correlation function is illustrated 
in Figure~\ref{fig:masses} (left). It shows the meson being 
created and destroyed by an operator $J$, which is implemented
when the quark propagators are tied together.   
At intermediate points the charm quark and antiquark interact 
with each other via the gluon fields and sea quarks in the background 
configuration. 

The meson mass is determined by fitting the average 
meson correlation function as a function of time on 
the lattice (we sum the end-points over $x$, $y$, $z$, at 
fixed $t$ to project onto zero spatial momentum for 
the meson). Because we are working with Euclidean time, 
the expected behaviour at large times 
is as an exponential (rather than the more normal 
phase factor):
\begin{equation}
\langle 0 | J^{\dag}(t_0+t) J(t_0) | 0 \rangle \equiv G(t) \stackrel{t \rightarrow \infty}{=} Ae^{-Mt} = Ae^{-Ma\times(t/a)}.
\label{eq:corr}
\end{equation}
The exponent is the mass of the lowest mass meson with 
the quantum numbers of the operator, $J$. 
The last piece of 
the equation above shows that, in fitting the correlation
function in terms of time on the lattice $t/a$ (i.e. 
the number of lattice spacings between two points in time) 
we will be able to determine the mass of the meson also 
in lattice units, i.e. the dimensionless combination $Ma$. We need 
to obtain a value for $a$ in order to convert this to 
physical, GeV, units. This is done by using another hadron mass 
(preferably one that is rather insensitive to quark masses) and 
setting the lattice result equal to the experimental value~\cite{pdg}. 
Quantities 
used for this include the radial excitation energy in the $\Upsilon$ 
system~\cite{Dowdallups} and the $\pi$ decay constant~\cite{DowdallpiK}.    

%%%%%%%%%%%%%%%%%%%%%%%%%%%%%%%%%%%%%%%%%%%%%%%%%%%%%%%%%%%%%%%%%%%%%%%%%
%%
%%   use this format to include an .eps figure into your paper
%%
\begin{figure}[t]
\centering
\includegraphics[width=7.0cm]{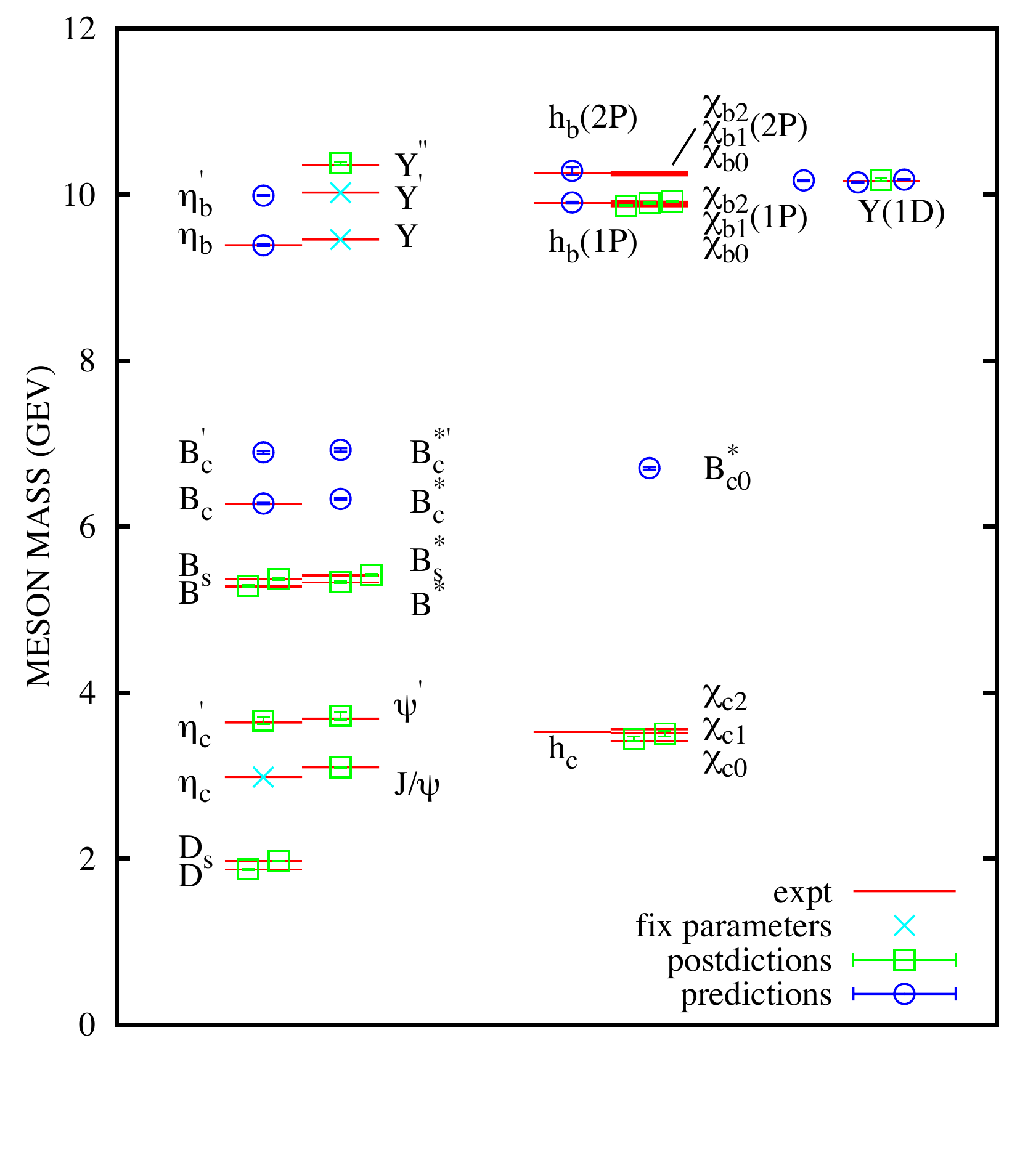}\hspace{1.0cm}
\includegraphics[width=7.0cm]{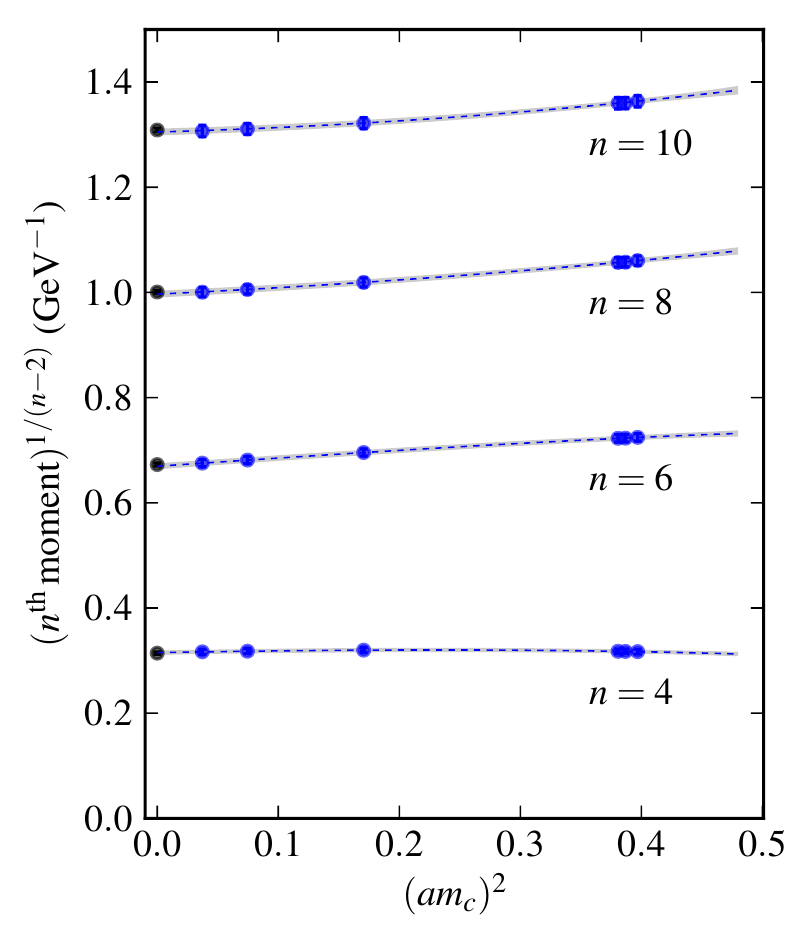}
\caption{Left: the gold-plated heavy meson spectrum from lattice QCD (points) 
compared to experiment~\cite{pdg} (red lines). Light cyan crosses denote those masses 
used to fix the parameters of QCD; green squares indicate postdictions and 
dark blue circles indicate predictions ahead of experiment. Recent lattice results 
are from~\cite{Dowdallups, Dowdallhl, Daldrop, Donaldpsi} Right: Moments 
of vector current-current correlators from lattice QCD plotted against 
the square of the lattice spacing~\cite{Donaldpsi}. The dashed line shows the continuum 
extrapolation. The black points at $a=0$ correspond to values extracted 
from experiment for the charm contribution to $R_{e^+e^-}$~\cite{karlsruhe2}. }
\label{fig:gold}
\end{figure}
%%%%%%%%%%%%%%%%%%%%%%%%%%%%%%%%%%%%%%%%%%%%%%%%%%%%%%%%%%%%%%%%%%%%%%%%%%%

Figure~\ref{fig:masses} (right) shows the correlator for the 
$c\overline{c}$ pseudoscalar meson, for which the 
lowest mass meson is the $\eta_c$. The quantity plotted 
is $G(t)e^{Mt}$, where $Ma$ is the value obtained 
for the ground-state mass from a fit to the correlator. This value 
is $Ma=1.32724(3)$ which corresponds to 2.982(3) GeV 
at this value of the lattice spacing 
($a=0.08784(9) \mathrm{fm} \equiv 1/2.2466(23) \,\mathrm{GeV}^{-1}$). 
This shows how accurately the $\eta_c$ mass can be obtained.  
The calculation required fixing the charm quark mass, also in lattice units. 
The value here, using the Highly Improved Staggered formalism~\cite{HISQ} 
for the $c$ quarks, was $m_ca=0.432$. The 0.1\% accuracy obtainable 
on the $\eta_c$ mass, means that the lattice $c$ quark mass (on 
which it is linearly dependent) can be tuned to a similar level 
of accuracy~\cite{fds}.  

Figure~\ref{fig:masses} also demonstrates the behaviour of the 
correlator. At large values of $t$ it is dominated by the 
ground-state $\eta_c$, so that $G(t)e^{Mt}$ is a constant. At 
shorter times this is not true. Then higher mass states (for 
example, radial excitations) contribute and their masses 
can be determined with a careful calculation (as discussed 
elsewhere in these Proceedings~\cite{prelcharm2013}). This region merges 
seamlessly with the region where the 
correlator is controlled by perturbative 
QCD. It is the short time region that we use to match the 
lattice $m_c$ to that in a continuum scheme,
as described in the next Section.    
  
The $\eta_c$ is only one of a range of meson masses that 
can be accurately determined from lattice QCD. Figure~\ref{fig:gold} 
shows a summary plot of the spectrum of `gold-plated' mesons containing $c$ and 
$b$ quarks from lattice QCD and its comparison with experiment. 
A gold-plated meson is one that has no strong Zweig-allowed decay mode and 
so has a very narrow width. 
The accuracy of many of these masses from lattice QCD is now 
at the few MeV level where 
we need to worry about and estimate electromagnetic effects missing from our 
pure QCD calculations~\cite{fds}. The agreement with experiment is excellent, 
providing a stringent test of QCD. Indeed, some of the masses were 
predicted ahead of experiment.    

Handling $c$ and $b$ quarks presents some difficulties in 
lattice QCD because they are relatively heavy. When the 
Dirac equation is discretised onto a lattice of points the 
covariant derivative is replaced by a finite difference and 
this is only correct up to systematic errors of $\mathcal{O}(a^2)$. 
The question is, what sets the scale for these errors? 
For hadrons made of light quarks, this will typically be 
the scale of QCD, i.e. a few hundred MeV. For heavy quarks 
it can be the quark mass itself. For $c$ quarks, $m_ca$ 
is around 0.4 for typical lattice spacing values of around 
0.1 fm. An error of $\mathcal{O}([m_ca]^2)$ could then be of 
size 20\%. Working with `improved' discretisations 
raises the power of $ma$ in the error and improves the situation. 
For the Highly Improved Staggered Quark (HISQ) action~\cite{HISQ} that is used here, 
the leading errors are $\alpha_s^2 (m_ca)^2$ and $(m_ca)^4$, which 
give errors of a few \% at $a=$ 0.1fm. It is important to obtain 
results at multiple values of the lattice spacing and extrapolate 
to $a=0$ to remove the discretisation errors. This extrapolation 
is relatively benign if a highly improved action is used and 
therefore the error in the final result from this extrapolation 
is small.

\section{The current-current correlator method}
\label{sec:currcurr}
\begin{figure}[t]
\centering
\includegraphics[width=12.0cm]{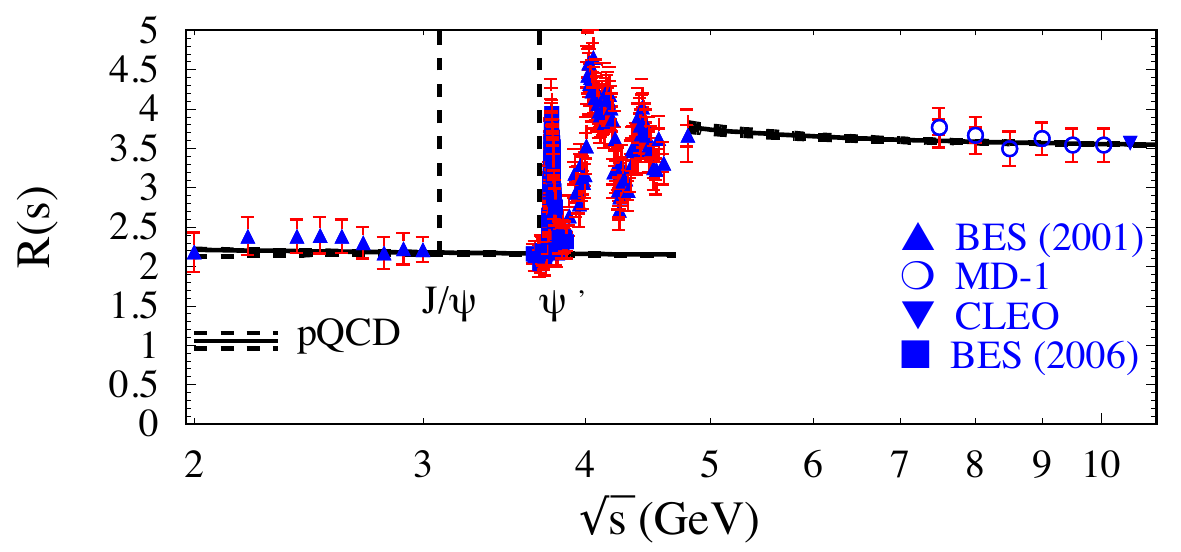}
\caption{$R_{e^+e^-}$ as a function of centre-of-mass energy, $\sqrt{s}$, 
around the charm threshold region. The solid line, with uncertainties given 
by the dashed lines, gives the prediction in perturbative QCD below and 
above the charm threshold. Figure from~\cite{karlsruhe2}.}
\label{fig:R}
\end{figure}

The production of a $c\overline{c}$ pair occurs directly in the real world 
in $e^+e^-$ collisions. Figure 1 (left) can also illustrate this case by 
representing the `heavy quark vacuum polarisation'. Then 
$J$ is the $c\overline{c}$ vector current which couples to the photon produced 
in $e^+e^-$. If we cut the diagram down the centre we expose a lot 
of quark-antiquark pairs and gluons produced from the original 
$c\overline{c}$ pair which, by unitarity, will end up as hadrons in 
the final state. Information about the charm quark vacuum polarisation 
can then be extracted from $\sigma(e^+e^- \rightarrow \mathrm{hadrons})$ 
if we can isolate the piece of the cross-section 
that corresponds to $c$ quark pair production. 
Because $R_{e^+e^-}=\sigma(e^+e^- \rightarrow \mathrm{hadrons})/\sigma_{\mathrm{point}}$ 
has step-like behaviour as a function of centre-of-mass energy $\sqrt{s}$ 
with well-separated heavy quark regions, this 
can be done using a mixture of theory and experiment. The contribution 
from $u$, $d$ and $s$ quarks can be calculated and subtracted, 
as illustrated in Figure~\ref{fig:R} from~\cite{karlsruhe2}.  
The basic tree-level 
QED calculation from textbooks~\cite{PeskinSchroeder} gives
$R_{e^+e^-}=3\sum_i Q^2_{q_i}$ for $i$ flavours 
of quarks with $Q_{q_i}$ the electric charge of that quark flavour in units of $e$, 
ignoring quark mass effects. 
QCD corrections can be incorporated that are impressively known up to and 
including $\alpha_s^3$ terms~\cite{Harlander}. The `natural' scale for $\alpha_s$ is $\sqrt{s}$, 
so this 
gives an accurate picture for $R$ in the region of a few GeV (below 
the charm threshold), and the agreement with experiment is good.  
Higher order electromagnetic contributions can be determined and they 
are very small; effects from the $Z$ are negligible.

%%%%%%%%%%%%%%%%%%%%%%%%%%%%%%%%%%%%%%%%%%%%%%%%%%%%%%%%%%%%%%%%%%%%%%%%%
%%
%%   use this format to include an .eps figure into your paper
%%
%%%%%%%%%%%%%%%%%%%%%%%%%%%%%%%%%%%%%%%%%%%%%%%%%%%%%%%%%%%%%%%%%%%%%%%%%%%

The $c$ quark contribution to $R_{e^+e^-}$, $R_c(s)$, then has pieces corresponding to 
the charm resonances (modelled as narrow peaks using the experimental 
information about each state), the charm threshold region 
(obtained from experiment after subtraction for $u$, $d$, and $s$ quarks) 
and the higher $s$ region above the charm threshold 
obtained from perturbation theory, again compared 
to experiment~\cite{karlsruhe2} or directly from experiment~\cite{Hoang}. 
Around the charm threshold region $R_c(s)$ is 
very sensitive to the charm quark mass and this can be used 
to determine $m_c$. 

The determination of $m_c$ uses analyticity properties 
to obtain the (dispersion) relationship 
between $s$-inverse moments of $R_c(s)$ and 
$q^2$-derivative 
moments of the charm quark vacuum polarisation function evaluated at 
$q^2=0$~\cite{PeskinSchroeder}: 
\begin{equation}
\mathcal{M}_{k,expt} \equiv \int \frac{ds}{s^{k+1}}R_c(s) = \mathcal{M}_{k,th} \equiv \left.\frac{12\pi^2}{k!}\left(\frac{d}{dq^2}\right)^k \Pi_c(q^2)\right|_{q^2=0}.
\label{eq:smom}
\end{equation}
$\mathcal{M}_{k,expt}$ is evaluated from $R_c(s)$ and 
the numbers are shown as the black points on the right-hand plot of 
Figure~\ref{fig:masses} (where $n = 2k+2$). Errors are 1\% or better.  
The contribution from the resonances dominates. 

$\mathcal{M}_{k,th}$, i.e. the $q^2$ derivatives of $\Pi_c$, needs 
evaluation of the behaviour of $\Pi_c$ at small $q^2$, i.e. 
in a very different kinematic region to that for $R_c(s)$. For heavy 
quarks, $q^2=0$ is well below the threshold to produce real 
quarks (so the $c$ quarks in Figure~\ref{fig:masses} (left) would be 
virtual). The expansion of $\Pi_c$ about $q^2=0$ can then be 
evaluated in QCD perturbation theory and the derivatives obtained, giving
for the vector current case
\begin{equation}
\mathcal{M}_{k,th} = Q_c^2 \frac{9}{4} C_{k,V} \left( \frac{1}{4m_c^2}\right)^k; \,\, C_{k,V} = C_{k,V}^{{0}} + \alpha_s C_{k,V}^{(1)} + \ldots .
\label{eq:moments}
\end{equation}
This exposes clearly the sensitivity of $\mathcal{M}_k$ to the $c$ quark 
mass. 
 $m_c$ in the equation above can be, for example, the $c$ quark mass in 
the $\overline{MS}$ scheme evaulated at the scale $\mu$. The perturbative 
series, $C_k$, is a power series expansion in $\alpha_s$. Its coefficients 
will reflect the scheme and scale chosen for $m_c$ so that the final 
result (to all orders) for $\mathcal{M}_k$ is scheme and scale invariant, and has the 
value obtained from experiment via $R_c$ in equation~\ref{eq:smom}. 
The `natural' scale for $\alpha_s$ here is  
$2m_c$, which is large enough for reasonably good control of the 
perturbative expansion.   

In fact, the QCD perturbation theory for $C_k$
has reached an extremely impressive level of calculation. 
Values for $C_k^{(3)}$ are known for the first few values of $k$, which 
corresponds to NNNLO~\cite{karlsruhepert, Boughezal, Maier}. Small values of $k$, 1 to 4, are preferred 
because larger values of $k$, although more sensitive to $m_c$, 
start to receive significant contributions from operators 
such as the gluon condensate 
which increase the uncertainty.  
Matching $\mathcal{M}_{k,th}$, with an input value of $\alpha_s$, 
to $\mathcal{M}_{k,expt}$ described above, Chetyrkin et al~\cite{karlsruhe2}
obtain, in the $\overline{MS}$ scheme with number of 
flavours, $n_f=4$,  $m_c(m_c)$ = 1.279(13) GeV. This 
is obtained from using the lowest moment, $k=1$, in equation~\ref{eq:moments}. 
The error is dominated by the experimental error in $R_c(s)$ and 
by the uncertainty taken in the value of $\alpha_s$ (3 times the current 
PDG uncertainty~\cite{pdg}).
The uncertainty coming from unknown higher order terms in the perturbation 
theory is estimated in the standard way by varying the scale, $\mu$, at which 
$\alpha_s$ is evaluated (varying the coefficients $C_{k,V}$ 
appropriately). The central value used here is 3 GeV, which is 
the same as the scale used for the central value of $m_c$ (subsequently 
iteratively run to the scale of $m_c$). The variation taken is $\pm$ 1 GeV~\cite{karlsruhe2}.  
Perturbative error estimates are always somewhat subjective 
and these have been criticised in~\cite{Hoang} as being too 
small, in particular pointing out that larger $\mu$ dependence can be seen 
when the $\mu$ in $\alpha_s$ and that in $m_c$ are decoupled.  

In the lattice QCD analysis described below we use the 
same perturbation theory but take a somewhat 
different approach to the perturbative errors, estimating directly 
the effect on $m_c$ of missing higher order terms 
using a Bayesian analysis. We are also able to fit multiple 
moments simultaneously and extract at the same time a value for $\alpha_s$.
These features improve the accuracy with which $m_c$ can be determined. 
They require the use of pseudoscalar current-current correlators
which are not accessible from experiment but, as discussed 
in Section~\ref{sec:lattice}, can be calculated very accurately in lattice QCD. 

%%%%%%%%%%%%%%%%%%%%%%%%%%%%%%%%%%%%%%%%%%%%%%%%%%%%%%%%%%%%%%%%%%%%%%%%%
%%
%%   use this format to include an .eps figure into your paper
%%
\begin{figure}[t]
\centering
\includegraphics[width=8.0cm]{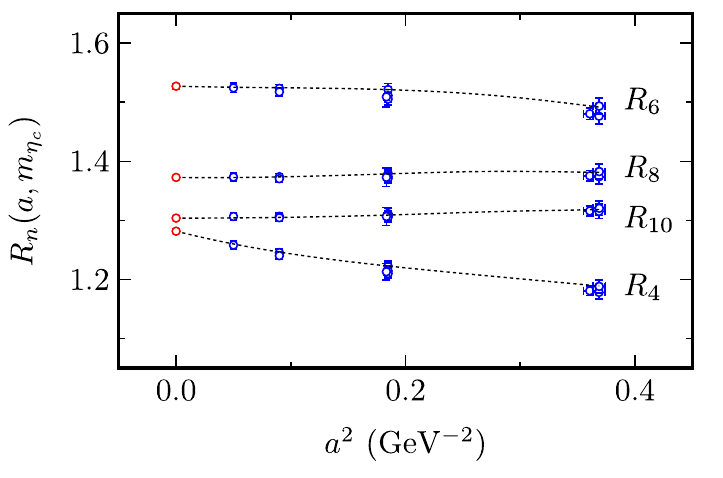}
\caption{Time moments of pseudoscalar $c\overline{c}$ correlators 
calculated in lattice QCD as a function of the square of the 
lattice spacing~\cite{hpqcd10}. The result extrapolated to $a=0$ can be used with 
continuum QCD perturbation theory to determine the $c$ quark 
mass in the $\overline{MS}$ scheme. }
\label{fig:rvasq}
\end{figure}
%%%%%%%%%%%%%%%%%%%%%%%%%%%%%%%%%%%%%%%%%%%%%%%%%%%%%%%%%%%%%%%%%%%%%%%%%%%

In lattice QCD we can substitute for $M_{k,expt}$ 
values of $\mathcal{M}_{k,latt}$ obtained
by taking time-moments 
of the $c\overline{c}$ meson correlation functions 
described in Section~\ref{sec:lattice}. 
The correlation functions (at zero spatial momentum) 
are the Fourier transform from 
energy to time-space of the charm 
quark vacuum polarisation function. Thus $q^2$-derivative 
moments become (squared) time-moments and we use~\cite{hpqcdkarlsruhe}
\begin{equation} 
G_n = \sum_t (t/a)^n G(t); \,\, n=2k+2 .
\end{equation}
$G(t)$ is a meson correlation function averaged over 
gluon field configurations, as in equation~\ref{eq:corr}. 
To compare to the continuum QCD 
perturbation theory of equation~\ref{eq:moments} we need to 
extrapolate $G_n$ to $a=0$ to obtain a continuum value. Thus 
we have to define $G(t)$ to be well-defined in that limit. For the 
HISQ formalism used here we have a PCAC 
relation (as in continuum QCD) that enables us to define an absolutely normalised 
pseudoscalar current operator: $J=(am_c) \overline{c}\gamma_5c$ and 
we use this to create and destroy pseudoscalar $c\overline{c}$ states 
in our correlation function. 

To perform the analysis~\cite{hpqcdkarlsruhe, hpqcd10} we calculate 
meson correlation functions at multiple values of the lattice 
spacing, fixing $am_c$ at each value of $a$ to be the value which 
gives the correct $\eta_c$ mass from the long-time behaviour of 
the correlator. We then calculate the time-moments as above for 
$n=$ 4, 6, 8 and 10 
(corresponding to $k=$ 1, 2, 3 and 4 in equation~\ref{eq:moments}).     
Taking time-moments emphasises the small, but non-zero, $t$ 
region of the correlation 
function since it is falling approximately exponentially (see Figure~\ref{fig:masses} right). 
Having a result for each moment at each value of $a$ then allows us 
to extrapolate to the continuum limit. 
The result for doing this for $c$ quarks is shown in Figure~\ref{fig:rvasq}.
To reduce discretisation errors we have actually plotted and extrapolated 
the ratio of $G_n$ to its value in the absence of gluon fields, $G_n^{(0)}$ 
(readily calculated by simply omitting the coupling to gluons 
in the Dirac equation). 
In fact we take: 
\begin{equation}
R_{n,latt} = G_4/G_4^{(0)},\,n=4;\,\, R_{n,latt} = \frac{aM_{\eta_c}}{2am_c}\left(\frac{G_n}{G_n^{(0)}}\right)^{1/(n-4)}, \,n=6,8,10 \ldots .
\label{eq:rdef}
\end{equation}
Here $aM_{\eta_c}$ and $am_c$ are lattice values. 
From Figure~\ref{fig:rvasq} we see that the extrapolation is 
relatively benign, especially as $n$ increases from 4. 
We include results from 4 values of 
the lattice spacing from 0.12 fm down to 0.045 fm. Values in 
the continuum limit have errors of order 0.1\%. 

At $a=0$ we can compare to the same ratio determined perturbatively:
\begin{equation}
R_{4,cont} = \frac{C_{1,PS}}{C_{1,PS}^{(0)}};\,\,R_{n,cont} = \frac{M_{\eta_c}}{2m_c(\mu)}\frac{C_{k,PS}}{C_{k,PS}^{(0)}};\,\, n=2k+2
\label{eq:rcont}
\end{equation}
where $C_{k,PS}$ is the full perturbative series for the pseudoscalar moment 
and $C_{k,PS}^{(0)}$ is the leading ($\alpha_s^0$) term. So $C_{k,PS}/C_{k,PS}^{(0)} = 1+c_1\alpha_s + \ldots$. For $n=4$ ($k=1$) 
in the pseudoscalar case we have no factor of masses in front of the series. 
This means that the $n=4$ moment is insensitive to the charm 
quark mass (it appears only 
in the scale for $\alpha_s$) and can be used to determine $\alpha_s$. 
The higher moments ($n=$ 6, 8, 10) can be used to determine $m_c(\mu)$ 
in terms of the (experimental) $\eta_c$ mass from equation~\ref{eq:rcont}.  

We match the lattice results at $a=0$ to the continuum perturbation theory, 
simultaneously fitting $n=$ 4, 6, 8 and 10 (including correlations 
between them and allowing for gluon condensate contributions) 
to extract $\alpha_s(\mu)$ 
and $M_{\eta_c}/m_c(\mu)$. The result we obtain for $m_c$ in the 
$\overline{MS}$ scheme with $n_f=4$ is $m_c(m_c)=1.273(6)$ GeV. Note 
that the calculation is done with 3 flavours of sea quarks and QCD 
perturbation theory is used to convert to 4 flavours. A complete 
error budget is given in~\cite{hpqcd10}. The error is dominated 
by the unknown higher order terms in the perturbative expansion of 
the moments and is 
estimated by including such terms with coefficients that are 
constrained by Bayesian priors. Information about the known $\mu$ 
dependence of the coefficients from the renormalisation 
group can be included this way. Our perturbative error is then 
about half the combined perturbative-$\alpha_s$ error in~\cite{karlsruhe1}. 
The statistical error from $\mathcal{M}_{k,latt}$ is much 
smaller in this case than that from $\mathcal{M}_{k,expt}$. 
The place in which experiment enters into the lattice 
calculation is in the tuning of the lattice $c$ quark 
mass using the experimental $\eta_c$ mass. For this we 
estimate the effect of missing electromagnetism from 
the lattice calculation; it is a tiny effect~\cite{fds}. 
There is no further error from missing electromagnetism 
because we are comparing a lattice QCD calculation to 
continuum QCD perturbation theory.   

A further test of this approach is to calculate the 
vector-vector correlator and compare the moments to those 
extracted from experiment via $R_c(s)$, 
$\mathcal{M}_{k,expt}$, described above~\cite{Donaldpsi}. 
To extrapolate the lattice vector charmonium correlator to 
$a=0$ we first have to renormalise the vector current. This 
we do using the continuum QCD perturbation theory for the 
$n=4$ ($k=1$) moment. Figure~\ref{fig:masses} (right) shows 
the comparison of lattice QCD vector moments against $a^2$ 
with the moments determined from experiment via $R_c(s)$ 
as the black points at $a=0$. The extrapolated lattice QCD 
results agree well with experiment, with the lattice QCD results 
having approximately double the error (including a small 
contribution allowing for higher order QED effects which are 
not included in the lattice calculation but are present in 
experiment). This then represents an impressive 1\% test of QCD 
and adds confidence to the determination of $m_c$ from 
the pseudoscalar moments. Using the lattice QCD 
vector moments to determine 
$m_c$ would give a result in agreement with that from 
the pseudoscalar but with a larger error. 
The long-time behaviour of the vector correlators simultaneously 
gives accurate results for the $J/\psi$ mass and leptonic width~\cite{Donaldpsi}.

\begin{figure}[t]
\centering
\includegraphics[width=7.5cm]{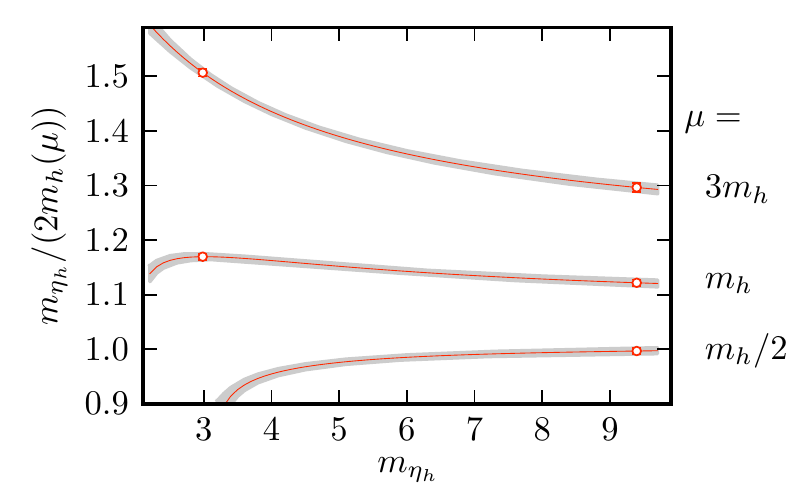}\hspace{0.5cm}
\includegraphics[width=7.0cm]{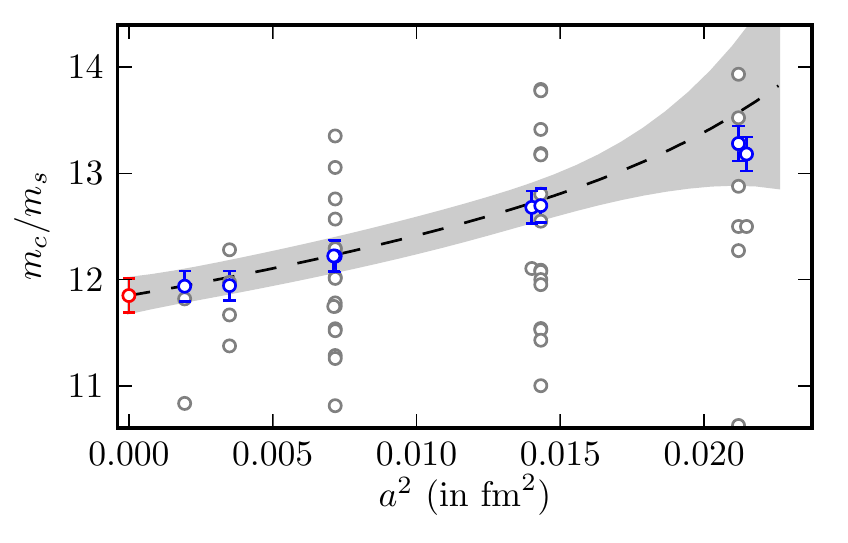}
\caption{Left: the ratio of pseudoscalar heavyonium mass to heavy quark 
mass in the $\overline{MS}$ scheme at scale $\mu$ as a function of heavyonium 
meson mass and for 3 different values of $\mu$~\cite{hpqcd10}. Notice how flat 
the curve is for ${m}_h({m}_h)$. Right: the ratio of 
$c$ to $s$ quark masses determined from lattice QCD plotted against the 
square of the lattice spacing. Extrapolation to $a=0$ gives 
the physical result 11.85(16)~\cite{mcms}.}
\label{fig:massrat}
\end{figure}

Because the HISQ action has small discretisation errors 
we can push to higher masses than $m_c$ and this 
was done in~\cite{hpqcd10}. By extrapolating up in mass 
we can also determine $m_b$ in the $\overline{MS}$ scheme 
from the same method:  
${m}^{(5)}_b({m}_b)=4.164(23) \mathrm{GeV}$.
Here the error is dominated by the extrapolation to the 
$b$ quark mass/$a=0$. 
The physical curve for the ratio of 
heavyonium mass to heavy quark mass is obtained on solving 
equation~\ref{eq:rcont},
and this is shown in Figure~\ref{fig:massrat} (left). 

%%%%%%%%%%%%%%%%%%%%%%%%%%%%%%%%%%%%%%%%%%%%%%%%%%%%%%%%%%%%%%%%%%%%%%%%%
%%
%%   use this format to include an .eps figure into your paper
%%
%%%%%%%%%%%%%%%%%%%%%%%%%%%%%%%%%%%%%%%%%%%%%%%%%%%%%%%%%%%%%%%%%%%%%%%%%%%

\section{Mass ratios}
\label{sec:massratios}
Lattice QCD enables us to determine the ratios of quark masses fully nonperturbatively. Provided that we have used the same lattice discretisation for both quarks, the 
$Z$ factors that connect the lattice quark mass to the $\overline{MS}$ quark 
mass at a given scale will cancel in the ratio. We then have, extrapolating to the continuum
\begin{equation}
\left(\frac{m_{q1,latt}}{m_{q2,latt}}\right)_{a=0} = \frac{m_{q1,\overline{MS}}(\mu)}{m_{q2,\overline{MS}}(\mu)}.
\end{equation}
Using the HISQ action for both $c$ and $s$ quarks, fixing $am_c$ 
from $M_{\eta_c}$ and $am_s$ from $M_K$ (via an unphysical $s\overline{s}$ 
pseudoscalar particle called the $\eta_s$), gives the results 
shown in Figure~\ref{fig:massrat} (right) as a function of lattice spacing. 
The extrapolated result, $m_c/m_s = 11.85(16)$ could not be obtained with 
this accuracy by any other method. It enables us to convert the accurate 
value for $m_c$ discussed in the previous Section into a 1\% determination 
of $m_s$. Running $m_s$ up from $m_c$ to the conventional 2 GeV gives 
$m_s(2\mathrm{GeV})$ = 92.2(1.3) MeV.  In~\cite{hpqcd10} we determine 
nonperturbatively the ratio of $m_b/m_c$, obtaining 4.51(4), and this 
acts as a check on the determination using moments and perturbation theory. 
Using mass ratios in this way we can leverage the accuracy in the 
heavy quark masses across the full set from $u$ to $b$~\cite{hpqcd10}. 
Amusingly we can use this to test the Georgi-Jarlskog expectation 
from GUTs that $m_b/m_s = 3m_{\tau}/m_{\mu}$~\cite{GeorgiJarlskog}. 
For $m_b/m_s$ we have  53.4(1.1) (allowing for some 
statistical correlation between 
$m_b/m_c$ and $m_c/m_s$). This is only in marginal agreement with 
$3m_{\tau}/m_{\mu}$ = 50.450(5)~\cite{pdg}. As lattice QCD 
calculations become more accurate there will be more tension in 
simple relationships of this kind, including those between quark 
masses and CKM elements~\cite{McNeile}.

\section{Conclusions}
\label{sec:conclusions}
Both continuum and lattice QCD determinations of the $\overline{MS}$ 
$c$ quark mass 
have reached a level of accuracy around 1\%. The most accurate 
result is ${m}_c^{(4)}({m}_c)=1.273(6)$ GeV using 
lattice QCD~\cite{hpqcd10}. This calculation used 3 flavours of sea quarks; 
future work is underway by both the ETM and HPQCD Collaborations to 
determine $m_c$ including $c$ quarks directly in the sea, thus removing 
any worries about the 3 to 4 flavour matching. To improve the accuracy 
on $m_c$ will be hard without having yet another order in QCD perturbation 
theory. It could be done by using a nonperturbative 
determination of $m_b/m_c$ and a more accurate result for $m_b$ from 
lattice QCD (using for finer lattices) because $m_b$ has a 
smaller perturbative error.  
It is important for phenomenologists to use these accurate values for 
quark masses in, for example, determination of Higgs cross-sections if 
they are to estimate reliably the uncertainty in the Standard Model 
cross-section. Currently the error on the value of $m_c$ being 
used~\cite{lhcwg1, lhcwg2, snowmasswg} is inflated by a factor of 3.  

\Acknowledgements
I am grateful to Steve King, Peter Lepage, Paul Mackenzie, Craig McNeile and 
Marcus Petschlies for 
useful discussions. Thanks also to the organisers for a very interesting meeting.

\end{document}